\documentclass[prl,showpacs,twocolumn,amsmath,amssymb,superscriptaddress]{revtex4-1}
\usepackage[colorlinks=true, citecolor=blue, urlcolor=blue ]{hyperref}
\usepackage{graphicx}
\usepackage{subfigure}
\usepackage{grffile}
\usepackage{bm}
\usepackage[version=3]{mhchem}
\usepackage{color}
\usepackage{hyperref}
\usepackage[normalem]{ulem}

\newcommand{\ket}[1]{{\vert #1\rangle}}

\newcommand{\braket}[2]{\langle#1\vert#2\rangle}
\newcommand{\eval}[3]{\langle#1\vert#2\vert#3\rangle}

\newcommand{\1}{\mbox{\bf 1}}
\newcommand{\ud}{\mathrm{d}}

\begin{document}

\title{Photoinduced phase switching from Mott insulator to metallic state in the quarter-filled Peierls-Hubbard model}

\author{Can Shao}
\email{shaocan@njust.edu.cn}
\affiliation{Department of Applied Physics $\&$ MIIT Key Laboratory of Semiconductor Microstructure and Quantum Sensing, Nanjing University of Science and Technology, Nanjing 210094, China}

\author{Takami Tohyama}
\email{tohyama@rs.tus.ac.jp}
\affiliation{Department of Applied Physics, Tokyo University of Science, Tokyo 125-8585, Japan}

\author{Hantao Lu}
\email{luht@lzu.edu.cn}
\affiliation{School of Physical Science and Technology, Lanzhou University, Lanzhou 730000, China}
\affiliation{Lanzhou Center for Theoretical Physics $\&$ Key Laboratory of Theoretical
Physics of Gansu Province, Lanzhou University, Lanzhou 730000, China}

\date{\today}

\begin{abstract}
Utilizing the exact diagonalization method, we investigate the one-dimensional Peierls-Hubbard model at quarter filling, where it manifests as an antiferromagnetic Mott insulator in units of dimers. By increasing the on-site Coulomb repulsion $U$, we observe a significant suppression of the Drude peak, based on a nonequilibrium linear response theory capable of capturing the zero-frequency (Drude) weight of the optical conductivity under periodic boundary conditions. However, after the ultrafast photoirradiation of this model with large $U$, we detect a distinct enhancement of the Drude peak, signifying the onset of a photoinduced insulator-metal transition. Comparing these dynamics with the half-filled Hubbard model and a noninteracting spinless half-filled Su-Schrieffer-Heeger model (corresponding to the quarter-filled Peierls-Hubbard model with infinite $U$), we propose a novel mechanism for the photoinduced metallic state: the empty-occupied and double-occupied dimers serve as the photoinduced charge carriers, akin to the holons and doublons in Hubbard model.
\end{abstract}

%\pacs{71.27.+a, 78.47.-p}

\maketitle

\section{Introduction}\label{sec1}
The exploration of nonequilibrium dynamics in strongly correlated systems opens up new avenues for disentangling the complicated interactions among different degrees of freedom and for realizing or manipulating quantum states, particularly within the realm of ultrafast pump-probe measurements~\cite{Murakami23}. Experimental observations have revealed photoinduced transitions from Mott insulators to metallic states in various materials, including the halogen-bridged Ni-chain compound~\cite{Iwai03}, the organic charge-transfer compound ET-F$_2$TCNQ~\cite{Okamoto07, Mitrano14}, and the undoped cuprates Nd$_2$CuO$_4$ and La$_2$CuO$_4$~\cite{Okamoto10}. In the underdoped cuprate compounds, phenomena such as the enhancement of superconductivity caused by microwave~\cite{Vedeneev08} and the emergence of light-induced superconducting-like phases~\cite{Fausti11,Nicoletti14,Hu14,Kaiser14} provide valuable insights into the origin of high-temperature superconductivity.

Theoretical investigations of photoinduced metallic state~\cite{Oka03, Oka05, Maeshima05, Takahashi08, Eckstein10,  Eckstein13, Shao16, Shinjo17, Werner19, Rincon21} and superconductivity~\cite{Werner18, Wang18, Bittner19, Werner19, Kaneko19, Wang21, Zhang23} in strongly correlated systems, mostly relying on half-filled Hubbard-type models, suggest that photoinduced charge carriers (holons and doublons) play a crucial role in generating the metallic state, while the pairing of these carriers contributes to the emergence of superconductivity. Recently, a photoinduced metallic state is studied in monoclinic VO$_2$ using the nonequilibrium cluster dynamical mean field
theory (cDMFT) and exact diagonalization (ED) methods. Their ED results, based on a Hubbard dimer, indicate that the mixed-orbital doublon state generated by laser pulses may contribute to the metallic state~\cite{Chen24}.

The one-dimensional (1D) Peierls-Hubbard model at quarter filling has also attracted significant attention because its ground state exhibits an antiferromagnetic Mott insulator in units of dimers~\cite{Le2020, Benthien2005}. The model can be used to describe a family of the charge-transfer salts~\cite{Pedron94, Nishimoto2000, Shibata01, Tsuchiizu01, Penc94, Mila95, Favand96, Benthien2005}. Photoinduced melting of charge order~\cite{Yonemitsu07, Yonemitsu_2009} and insulator-metal transition~\cite{Lee09} have been reported by incorporating Holstein types of electron-phonon couplings to the Peierls-Hubbard model, aligning with the photoresponse in corresponding materials like (EDO-TTF)$_2$PF$_6$~\cite{Matthieu05}. Furthermore, the ultrafast dynamics of the 1D extended Peierls-Hubbard model (including additional nearest-neighbour interactions) has been investigated, with a focus on the photoexcitation of electronic instabilities and polarization-inverted domains~\cite{Yamaguchi19, Rincon14}.

In this paper, we propose a novel mechanism for the photoinduced metallic state in the 1D quarter-filled Peierls-Hubbard model, based on the ED method. In equilibrium, increasing the interaction strength in a dimerized chain results in a significant suppression of the Drude weight in the optical conductivity, indicating the formation of an insulating state. Upon ultrafast photoirradiation of the system, we observe a substantial enhancement of the Drude peak, suggesting a photoinduced insulator-metal transition. Unlike the scenario in the half-filled Hubbard model where photoinduced charge carriers such as holons and doublons contribute to the appearance of Drude peak, we propose that the empty-occupied and double-occupied dimers serve as the photoinduced charge carriers in the quarter-filled Peierls-Hubbard model, thus facilitating the transition to a metallic state. %\shao{To some extent, this feature is analogous to the mixed-orbital doublon state studied in monoclinic VO$_2$~\cite{Chen24}, but in a single-orbital context.}

\section{Models and measurements}\label{sec_model}

The 1D Peierls-Hubbard model can be written as $H=H_k+H_U$, where the kinetic part
\begin{eqnarray}
H_k=-\sum_{i,\sigma}\left(t_h(1+\delta(-1)^{i}) c^{\dagger}_{i,\sigma} c_{i+1,\sigma}+\text{H.c.}\right)
\label{H1}
\end{eqnarray}
and the interaction term
\begin{eqnarray}
H_U=U\sum_{i}n_{i,\uparrow}n_{i,\downarrow}.
\label{H2}
\end{eqnarray}
Here $c^{\dagger}_{i,\sigma}$ ($c_{i,\sigma}$) is the creation (annihilation) operator of an electron at site $i$ with spin $\sigma$, and $n_{i,\sigma}$ the number operator. $t_h(1+\delta(-1)^{i})$ denotes the staggered hopping constants and we set $\delta=0.5$ throughout the paper. We consider a periodic chain with $L$ sites and $U$ represents the strength of on-site Coulomb interaction. $L$ is set to be $12$ for the 1D Peierls-Hubbard model, and it is important to note that the number of the unit cells is $L/2=6$. In this paper, we use units with $e=\hbar=c=a_0=1$, where $e$, $\hbar$ and $c$ are the elementary charge, the reduced Planck constant and the speed of light, respectively, and $2a_0$ is the lattice constant. With these units, $t_h$ and ${t_h}^{-1}$ are the unit of energy and time, respectively. Our focus is on zero temperature, with the system evolving from its ground state.

We utilize the time-dependent optical conductivity to characterize the photoinduced metallic behavior. The external electric field of a pumping pulse is included via the Peierls substitution
\begin{equation}
c^{\dagger}_{i,\sigma}c_{i+1,\sigma}+\text{H.c.}\rightarrow
e^{\mathrm{i}A(t)}c^{\dagger}_{i,\sigma}c_{i+1,\sigma}+\text{H.c.}.
\label{eq:Peierls}
\end{equation}
$A(t)$ represents the vector potential of the pumping pulse, which can be expressed as
\begin{equation}
A(t)=A_0e^{-\left(t-t_0\right)^2/2t_d^2}\cos\left[\omega_0\left(t-t_0\right)\right].
\label{eq:vpotent}
\end{equation}
This is a cosine function with a Gaussian envelope centered at $t_0$. The parameters $t_d$ and $\omega_0$ control its width and central frequency, respectively. To obtain the time-dependent wave function, the Lanczos method~\cite{Prelovsek} is employed with the key formula
% \begin{eqnarray}
\begin{equation}
\ket{\psi(t+\delta{t})}=e^{-\mathrm{i}H(t)\delta t}\ket{\psi(t)}
\simeq\sum_{l=1}^{M}{e^{-\mathrm{i}\epsilon_l\delta{t}}}\ket{\phi_l}\braket{\phi_l}{\psi(t)},
\label{eq:lanczos}
\end{equation}
% \end{eqnarray}
where $\epsilon_l$ and $|\phi_l\rangle$ are the Krylov eigenvalues and eigenvectors generated in the Lanczos process, respectively. Note that the initial state is the ground state of Hamiltonian before pump. In our simulations, we choose the time step $\delta{t}=0.02$ and  $M=30$ to ensure the convergence of numerical results.

The time-dependent optical conductivity $\sigma(\omega,t)$ can be calculated by the nonequilibrium linear-response theory~\cite{Zala14}:
\begin{equation}
\sigma(\omega,t)=\int_0^{+\infty}\sigma(t+s,t)e^{\mathrm{i}(\omega+i\eta)s}\,\ud s,
\label{eq:sigmaomega}
\end{equation}
where the response function $\sigma(t',t)$ (with $t'\ge t$) reads
\begin{equation}
\sigma(t',t)=\frac{1}{L}\left[\eval{\psi(t')}{\tau}{\psi(t')}+\int_t^{t'}\chi(t',t'')\,\ud t''\right].
\label{eq:sigmatt}
\end{equation}
The first term of Eq.~(\ref{eq:sigmatt}) is referred to as the diamagnetic term, and the 1D stress tensor operator reads
\begin{equation}
\tau=\sum_{i,\sigma}\left(t_h e^{\mathrm{i}A(t)}(1+\delta(-1)^{i}) c^{\dagger}_{i,\sigma} c_{i+1,\sigma}+\text{H.c.}\right).
\label{eq:tau}
\end{equation}
In the second term of Eq.~(\ref{eq:sigmatt}), the two-time susceptibility
\begin{equation}
\chi(t',t'')=-\mathrm{i}\theta(t'-t'')\eval{\psi(t)}{[j^{I}(t'),j^{I}(t'')]}{\psi(t)},
\label{eq:chi}
\end{equation}
where $j^{I}(t')=U^\dagger(t',t)\,j\,U(t',t)$ is the interaction representation of the current operator, with $U(t',t)$ the time-evolution operator and the current operator
\begin{eqnarray}
j=-\mathrm{i}\sum_{i,\sigma}\left(t_h e^{\mathrm{i}A(t)}(1+\delta(-1)^{i}) c^{\dagger}_{i,\sigma} c_{i+1,\sigma}-\text{H.c.}\right).
\label{j}
\end{eqnarray}
Note that in Eq.~(\ref{eq:sigmaomega}), the integration cutoff for $s$ is taken to be one hundred time units, and the broadening factor $\eta=0.1$.

In what follows, we study the real part of the optical conductivity, i.e., Re $\sigma(\omega,t)$, before and after pump.
The detailed derivation of the time-dependent optical conductivity can be found in Ref.~\cite{Zala14}, and its validity has been discussed in Ref.~\cite{Shao16}.

\section{Results}\label{sec_pump}

\begin{figure}
\centering
\includegraphics[width=0.5\textwidth]{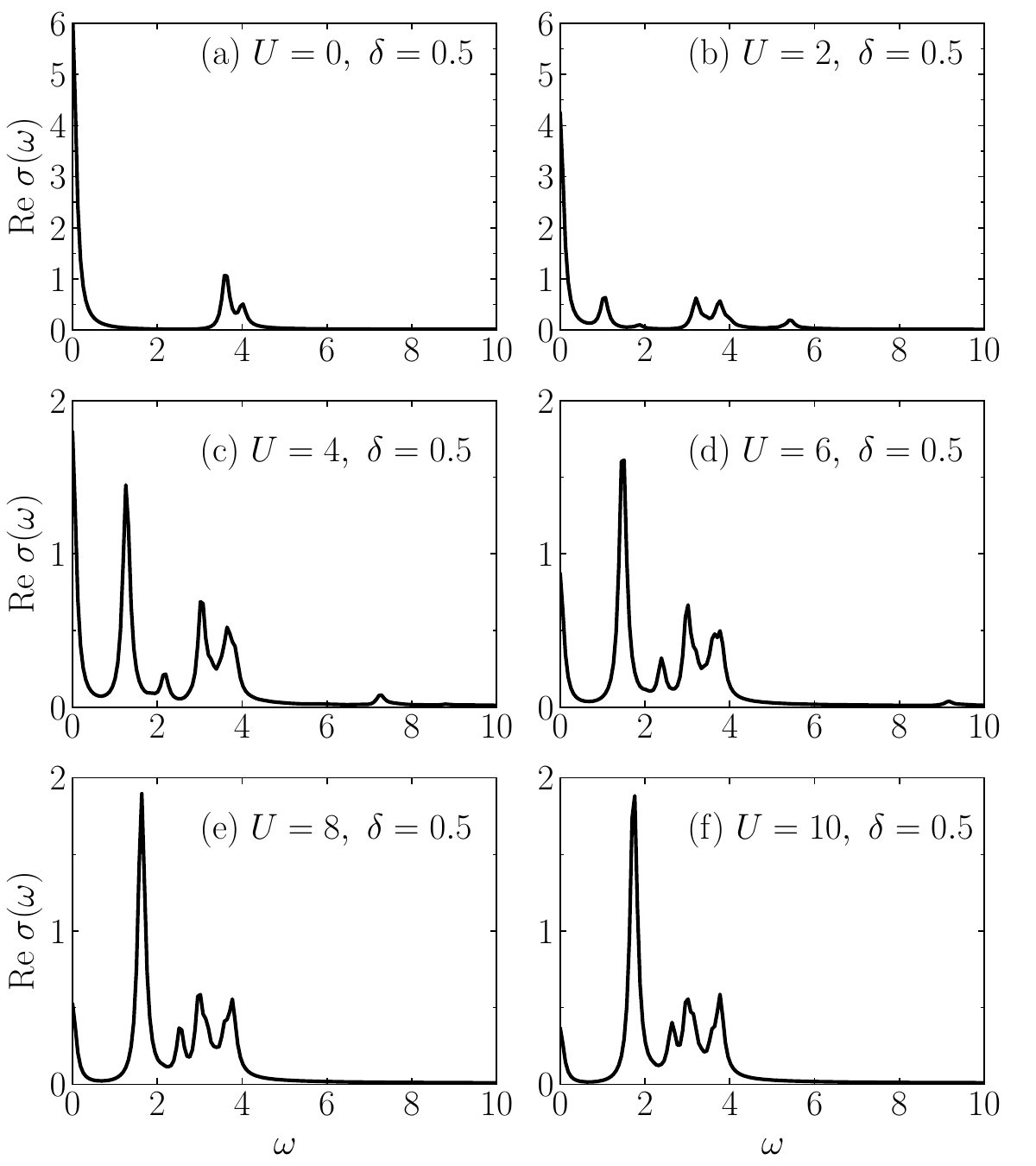}
\caption{(Color online) The optical conductivity Re $\sigma(\omega)$ in equilibrium for the quarter-filled Peierls-Hubbard with $L=12$, $\delta=0.5$, and (a) $U=0.0$, (b) $U=2.0$, (c) $U=4.0$, (d) $U=6.0$, (e) $U=8.0$, and (f) $U=10.0$, respectively.}
\label{fig_equilibrium2}
\end{figure}

Results of the optical conductivity before the pump (in equilibrium), Re $\sigma(\omega)$, are depicted in Fig.~\ref{fig_equilibrium2} with different values of $U$. Note that the optical conductivity obtained from the nonequilibrium linear-response theory~\cite{Zala14} provides insights into the Drude weight, i.e., the zero-frequency weight of optical conductivity. For $U=0$ in Fig.~\ref{fig_equilibrium2}(a), a prominent Drude peak can be observed along with additional structures within the range of $\omega\in[3,4.5]$.
As $U$ increases from $0$ to $4$, as shown in Figs.~\ref{fig_equilibrium2}(a), \ref{fig_equilibrium2}(b) and \ref{fig_equilibrium2}(c), the Drude peak lose its weight, while a side peak around $\omega=1.0$ emerges and grows. It is worth noting that the $y$-axis range in Figs.~\ref{fig_equilibrium2}(a) and \ref{fig_equilibrium2}(b) is $[0,6]$, whereas in Figs.~\ref{fig_equilibrium2}(c), \ref{fig_equilibrium2}(d), \ref{fig_equilibrium2}(e) and \ref{fig_equilibrium2}(f) it is $[0,2]$. This side peak becomes the main peak when $U\geq6$, and its position shifts to the right to $\omega=1.76$ when $U=10$, connecting to the left-shifted structures originally located at $\omega\in[3,4.5]$ in Fig.~\ref{fig_equilibrium2}(a). Despite a small remaining Drude peak in Fig.~\ref{fig_equilibrium2}(f) with $U=10$, attributed to the finite-size effect in PBC even for the insulator state~\cite{Shao16}, the nonequilibrium linear-response theory proves useful in characterizing the metallic state in and out of equilibrium by analyzing the relative magnitudes of Drude peak. In addition, our results of equilibrium optical conductivity based on the ED method are similar to the results from the density-matrix renormalization group method~\cite{Benthien2005}, despite slight variations in model parameters.

\begin{figure}[t]
\includegraphics[width=0.5\textwidth]{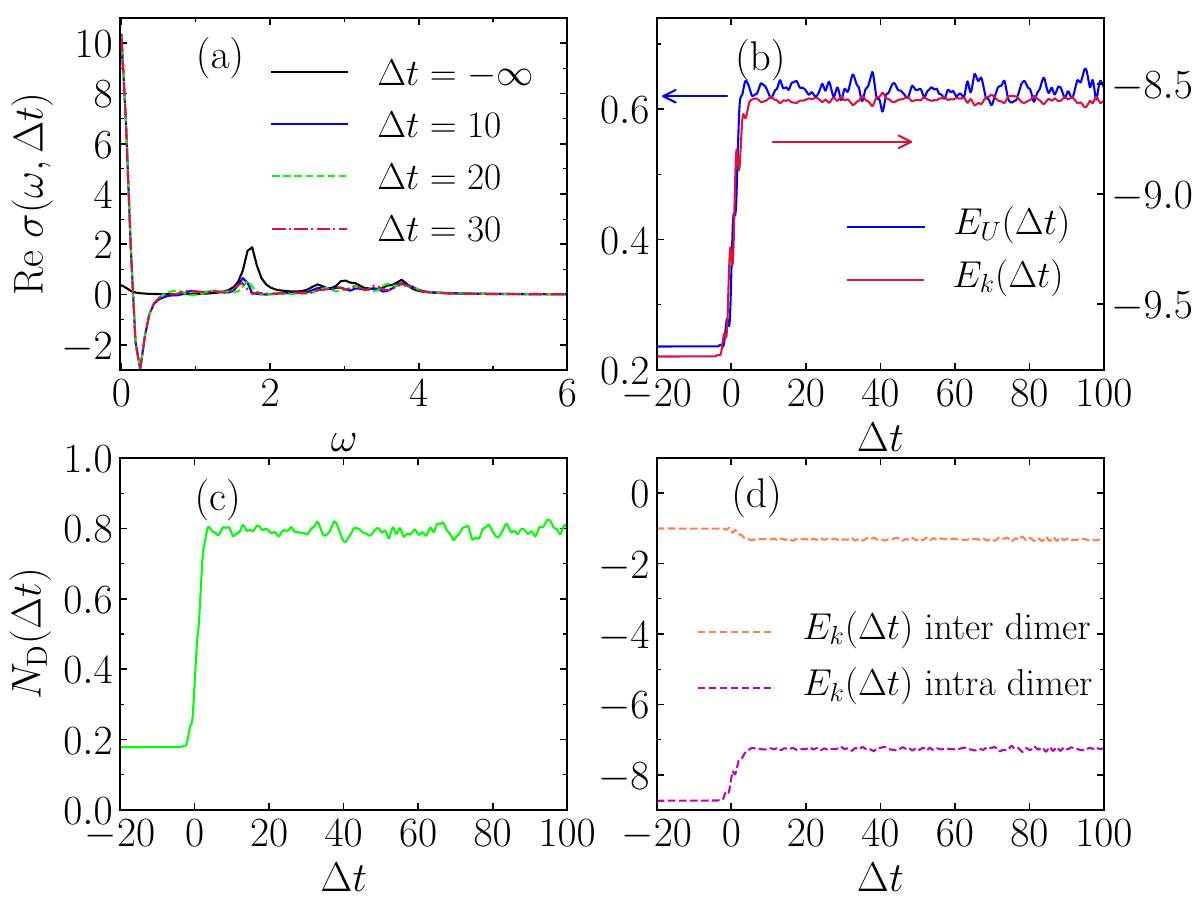}
\caption{(Color online) (a) The time-dependent optical conductivity Re~$\sigma(\omega, \Delta t)$, (b) the time-dependent kinetic energy $E_k(\Delta t)$ and interaction energy $E_U(\Delta t)$, (c) the number of the double-occupied dimers $N_{\rm D}$ as a function of time, and (d) the time-dependent inter- and intra-dimer kinetic energys. Parameters of the quarter-filled Peierls-Hubbard model: $L=12$, $\delta=0.5$ and $U=10.0$. Parameters of the pumping pulse: $A_0 = 0.2$, $t_d = 2.0$, $\omega_0=1.76$.}
\label{fig_pump}
\end{figure}

In the nonequilibrium scenario, we consider the quarter-filled Peierls-Hubbard model with $L=12$, $\delta=0.5$ and $U=10.0$, which is driven by a laser pulse described by Eq.~(\ref{eq:vpotent}). The laser frequency $\omega_0=1.76$ matches the main peak of the equilibrium optical conductivity. In the following discussions, we set the central time of pumping pulse $t_0=0$ and substitute the variable $t$ in $\sigma(\omega,t)$ with $\Delta t=t-t_0$, representing the time difference between the probing time and $t_0$. The real part of the optical conductivity, Re $\sigma(\omega, \Delta t)$, for $\Delta t=-\infty$, $10$, $20$ and $30$ is depicted in Fig.~\ref{fig_pump}(a) with different colors, where $\Delta t=-\infty$ indicates the equilibrium optical conductivity without the application of the pumping pulse. We observe the emergence of a dominant peak at $\omega=0$ after photoirradiation, signifying a photoinduced metallic state. To analyze the details of the nonequilibrium dynamics, we present the time-dependent kinetic energy $E_k(\Delta t)$, which is composed of the intra- and inter-dimer kinetic energies, and the time-dependent interaction energy $E_U(\Delta t)$ in Fig.~\ref{fig_pump} (b). We find that $E_U$ increases from $0.24$ to $0.63$ after the pumping pulse. Referring to Eq.~(\ref{H2}) where $E_U=U n_d$, $n_d$ represents the total doublon number and $U=10$ in the nonequilibrium discussions. It is worth noting that the doublon number is counted over sites, rather than dimers. As a result, the increase of doublon number caused by the pumping pulse is approximately $0.04$ on the $12$-site lattice. This suggests that the photoinduced doublons are insufficient to produce a metallic state.

However, the increase of kinetic energy amounts to $1.1$, indicating that more than $70$ percent of the injected energy contributes to the kinetic energy. Furthermore, from Fig.~\ref{fig_pump} (d), we observe that the intra-dimer kinetic energy increases, while the inter-dimer kinetic energy decreases. It is important to note that in equilibrium (before pump), the intra-dimer kinetic energy is smaller than the inter-dimer kinetic energy due to the formation of the hybridization gap (bonding-antibonding separation). Specifically, every dimer is occupied only by one electron, forming the antiferromagnetic ground state in units of dimers. With the injection of energy from the pump, electrons are facilitated in overcoming the barrier among dimers, leading to the appearance of empty-occupied and double-occupied dimers. The double-occupied dimer with one spin-$\uparrow$ electron and one spin-$\downarrow$ electron has four configurations: ($\uparrow$,$\downarrow$), ($\downarrow$,$\uparrow$), ($\downarrow$$\uparrow$,o) and (o,$\downarrow$$\uparrow$). Here o represents an empty site in the dimer. We present the number of the double-occupied dimers, $N_{\rm D}$, as a function of time in Fig.~\ref{fig_pump} (c), where we observe that its increment is more than $0.6$ for our $6$-dimer chain. We thus propose that the empty-occupied and double-occupied dimers serve as the photoinduced carriers contributing to a metallic state. Additionally, both empty-occupied and double-occupied dimers can result in less mobile electrons within dimers and more mobile electrons between dimers. This leads to an increase of the intra-dimer kinetic energy and a decrease of the intra-dimer kinetic energy (since kinetic energy is zero when there is no motion), as shown in Fig.~\ref{fig_pump} (d).

\begin{figure}[t]
\includegraphics[width=0.5\textwidth]{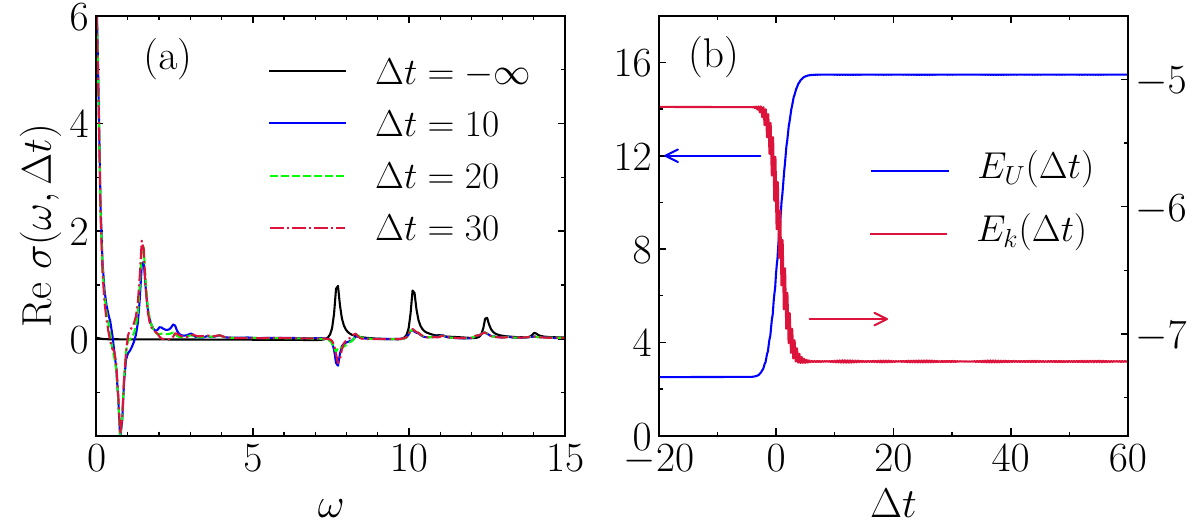}
\caption{(Color online) (a) The time-dependent optical conductivity Re $\sigma(\omega, \Delta t)$, (b) the time-dependent kinetic energy $E_k(\Delta t)$ and interaction energy $E_U(\Delta t)$, (c) the time-dependent current $j(\Delta t)$, and (d) the time-dependent order parameter of spin-density-wave order $\mathcal{O}_{\rm SDW}(\Delta t)$. Parameters of the half-filled Hubbard model: $L=10$ and $U=10.0$. Parameters of the pumping pulse: $A_0 = 0.2$, $t_d = 2.0$, $\omega_0=7.68$.}
\label{fig_3}
\end{figure}

This phenomenon is reminiscent of the photoinduced carriers, i.e., holons or doublons, in the half-filled Hubbard model. Especially, the low-energy negative spectral weight in the postpump optical conductivity shown in Fig.~\ref{fig_pump} (a) can also be observed in the half-filled Hubbard model, see Fig.~\ref{fig_3}(a) which we will discuss later. As also mentioned in Ref.~\cite{Lu15}, this negative peak at low frequency originates from an excitation from the optically allowed odd-parity state to an optically forbidden even-parity state. Furthermore, the energy separation between them vanishes when either $U$ or $L$ becomes infinite, indicating that this negative peak can be attributed to the finite size effect. More details on the analysis of the energy separation between the odd- and even-parity states can be found in the Appendix.

Now we discuss the photoinduced metallic state in the 1D half-filled Hubbard model in order to compare the results with those of the 1D quarter-filled Peierls-Hubbard model. Here we set the lattice size $L=10$ and the interaction $U=10$, respectively. The equilibrium optical conductivity is shown in Fig.~\ref{fig_3}(a) with a black line, where the primary absorbing peak is observed at $\omega=7.68$. Setting the central frequency of the pumping pulse, $\omega_0$, to $7.68$ and choosing $A_0=0.2$ and $t_d=2.0$ to be consistent with the laser parameters used in the quarter-filled Peierls-Hubbard model, we observe the emergence of the zero-frequency Drude peak and a low-frequency negative peak in the time-dependent optical conductivity, Re $\sigma(\omega, \Delta t)$, with $\Delta t=10$, $20$ and $30$, as shown in Fig.~\ref{fig_3}(a) after the pumping pulse. The presence of the Drude peak signifies the onset of a photoinduced metallic state, while the emergence of the low-frequency negative peak is associated with the excitation between the odd- and even-parity states mentioned earlier. With an increment in interaction energy of $13$, corresponding to an increase in doublon number by $1.3$ for a $10$-site lattice, the kinetic energy decreases from $-5.2$ to $-7.2$. This decrease suggests an increase of the mobility of carriers (doublons), including the formation of $\eta$ pairing discussed in Ref.~\cite{Kaneko19}. It becomes evident that photoinduced carriers and their transport contribute to the formation of photoinduced metallic state. This behavior is very similar to the results observed in the 1D quarter-filled Peierls-Hubbard model, where the photoinduced carriers are empty-occupied and double-occupied dimers instead.

\begin{figure}[t]
\includegraphics[width=0.5\textwidth]{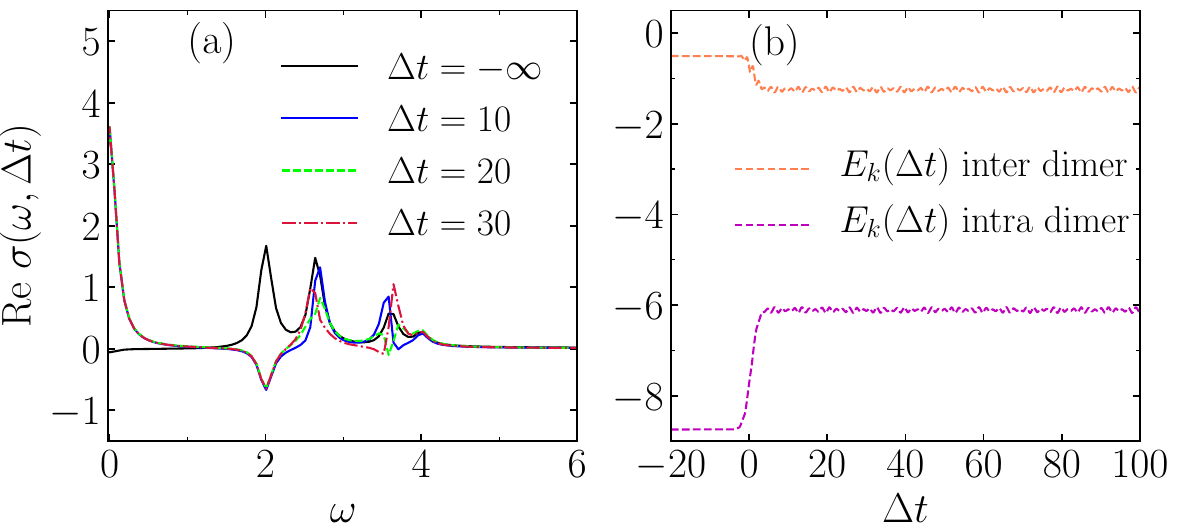}
\caption{(Color online) (a) The time-dependent optical conductivity Re $\sigma(\omega, \Delta t)$, (b) the time-dependent kinetic energy $E_k(\Delta t)$ inter- and intra-dimers. Parameters of the noninteracting hall-filled SSH model: $L=12$ and $\delta=0.5$. Parameters of the pumping pulse: $A_0 = 0.2$, $t_d = 2.0$, $\omega_0=2.00$.}
\label{fig_4}
\end{figure}

The photoinduced metallic behavior of a noninteracting half-filled Su-Schrieffer-Heeger (SSH) model is also evident in the time-dependent optical conductivity in Fig.~\ref{fig_4}(a). The SSH model in our paper is based on a $6$-dimer chain with six spinless electrons filled in. Comparatively, the quarter-filled Peierls-Hubbard model is also based on a $6$-dimer chain, but with three spin-$\uparrow$ and three spin-$\downarrow$ electrons. When the interaction strength $U$ approaches infinity, the quarter-filled Peierls-Hubbard model transforms into this spinless half-filled SSH model. We examine their different mechanisms for photoinduced metallic state. The central frequency of the pumping pulse, $\omega_0=2.0$, aligns with the first absorbing peak of the optical conductivity before pump. We set $A_0=0.2$ and $t_d=2.0$, consistent with the laser parameters applied on the quarter-filled Peierls-Hubbard model. After the pump, the first absorbing peak of the optical conductivity loses its weight and the Drude peak emerges, as shown in the time-dependent optical conductivity, Re $\sigma(\omega, \Delta t)$ with $\Delta t=10$, $20$ and $30$, in Fig.~\ref{fig_4}(a). While, the absence of a low-energy negative peak in the time-dependent optical conductivity confirms our results in Appendix, indicating that as $U$ becomes infinite in the quarter-filled Peierls-Hubbard model, the energy separation between the odd- and even-parity states vanishes, and negative low-frequency peak will not emerge.
On the other hand, the noticeable increase (decrease) observed in the time-dependent kinetic energy intra-(inter-)dimers, as depicted in Fig.~\ref{fig_4}(b), compared to those shown in Fig.~\ref{fig_pump} (d) in the quarter-filled Peierls-Hubbard model, can be attributed to the limited mobility of photoinduced double-occupied electrons within dimers, stemming from their spinless fermionic nature.

\section{Summary and discussion}\label{sec_conclusion}

To summarize, employing the exact diagonalization method, we presented the equilibrium outcomes of the optical conductivity in the 1D quarter-filled Peierls-Hubbard model. Starting with a purely dimerized chain, we observe that increasing $U$ leads to the suppression of the Drude weight in the optical conductivity. This confirms previous studies suggesting that this model manifests an antiferromagnetic insulator, arising from the combined effects of Peierls instability and electron correlations. Under nonequilibrium conditions, we observed the emergence of a photoinduced metallic state following a pumping pulse, as indicated by the appearance of a prominent Drude peak. However, we noted that the increase of photoinduced carriers, i.e., doublon numbers, is very small compared to the photoinduced metallic state in the pure Hubbard chain at half-filling. Instead, we propose that the empty-occupied and double-occupied dimers can serve as the photoinduced carriers contributing to the transition from a Mott insulator to a metallic state.

% Therefore, we suggest a high-time-resolution trARPES measurement on the relevant CDW insulators, such as the Br-Bridged Pd compounds~\cite{Sasaki07}, or on ultracold atoms platforms. This can deepen our current understanding from the dynamic perspective of the intermediate BOW state caused by the competition of many-body effects.

\begin{acknowledgments}
C. S. acknowledges support from the National Natural Science Foundation of China (NSFC; Grant No. 12104229) and the Fundamental Research Funds for the Central Universities (Grant No. 30922010803).
T. T. is partly supported by the Japan Society for the Promotion of Science, KAKENHI (Grant No. 24K00560) from the Ministry of Education, Culture, Sports, Science, and Technology, Japan.
H. L. acknowledges support from the National Natural Science Foundation of China (NSFC; Grants No. 11474136 and No. 12247101).
\end{acknowledgments}

\appendix*

\section{Appendix\\ Analysis of the energy separation between the odd- and even-parity states}\label{app}

\begin{figure}[t]
\centering
\includegraphics[width=0.45\textwidth]{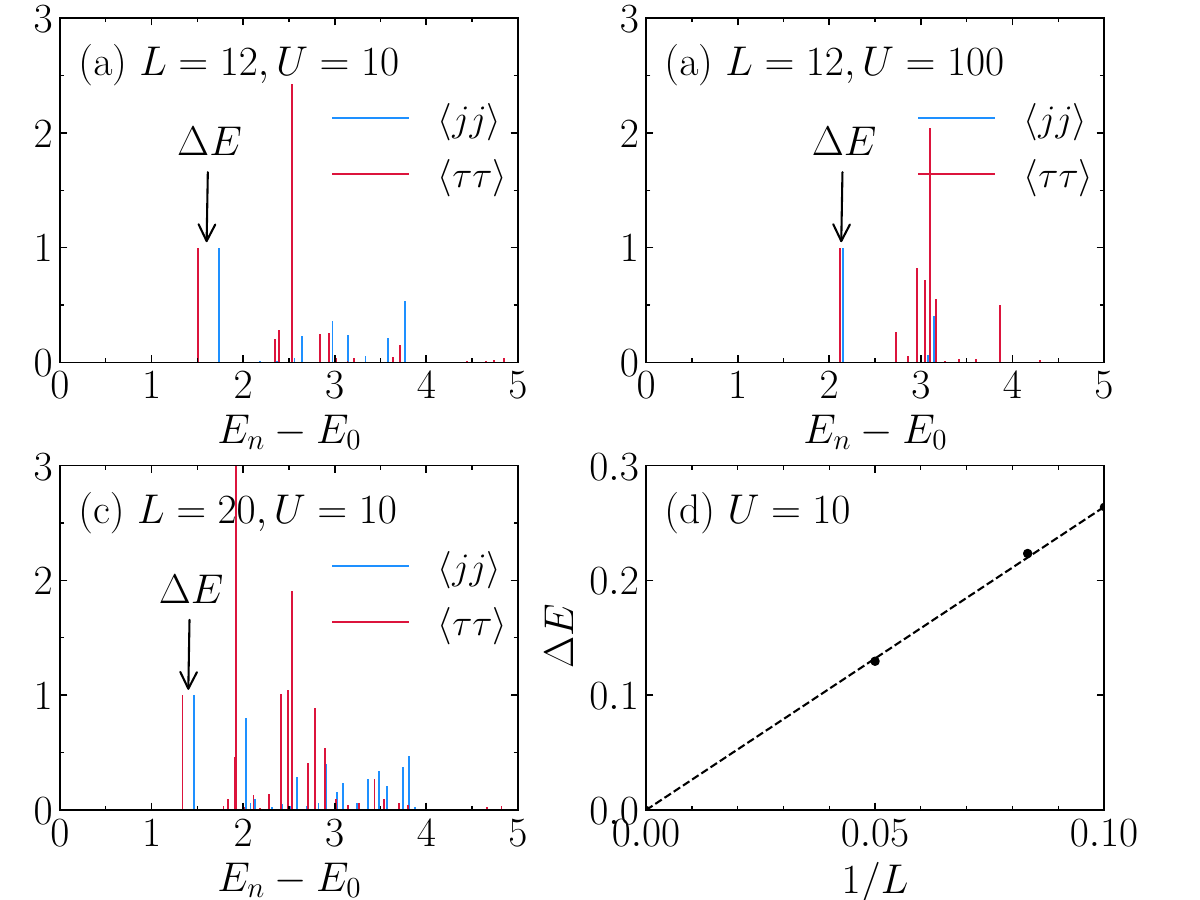}
\caption{The correlation function $\langle j j \rangle$ (blue lines) and $\langle \tau \tau \rangle$ (red lines) with $L=12$ and $U=10$ in (a), $L=12$ and $U=100$ in (b), and $L=20$ and $U=10$ in (c), respectively. (d) The energy separation between the odd- and even-parity states, $\Delta E$, as a function of $1/L$.}
\label{fig_S1}
\end{figure}

%\section{(I) Analysis of the energy separation between the odd- and even-parity states}\label{SecI}

In the main text, after photoirradiation of the one-dimensional (1D) quarter-filled Peierls-Hubbard model, we observe the emergence of a low-energy negative peak at $\omega\approx0.23$ in the postpump optical conductivity. Similar to observations in the 1D Hubbard model~\cite{Lu15}, this negative peak is attributed to an excitation from the optically allowed odd-parity state to an optically forbidden even-parity state. To confirm this, we calculate the current-current correlation function
\begin{equation}
\langle j j \rangle=\frac{\pi}{L}\sum_a |\langle a|j|0\rangle|^2\delta(\omega-E_a+E_0),
\label{eq:lanczos}
\end{equation}
and the correlation function replacing the current operator $j$ with the stress tensor operator $\tau$
\begin{equation}
\langle \tau \tau \rangle=\frac{\pi}{L}\sum_b |\langle b|\tau|0\rangle|^2\delta(\omega-E_b+E_0).
\label{eq:lanczos}
\end{equation}
Considering the symmetries of the operators $j$ and $\tau$ (see their definition the main text), $|a\rangle$ and $|b\rangle$ must be the odd- and even-parity states, respectively~\cite{Makoto02}.

We present the results of $\langle j j \rangle$ (blue lines) and $\langle \tau \tau \rangle$ (red lines) as a function of $E_n-E_0$ in Fig.~\ref{fig_S1} (a) with $L=12$ and $U=10$, in Fig.~\ref{fig_S1} (b) with $L=12$ and $U=100$, in Fig.~\ref{fig_S1} (c) with $L=20$ and $U=10$, respectively. Notice that $L$ and $U$ represent the lattice size and the interaction strength, respectively, and the results are normalized by odd and even state we can concerned with. In Fig.~\ref{fig_S1} (a), the normalized delta peak of $\langle j j \rangle$ corresponds to the main peak of the equilibrium optical conductivity, as seen in Fig.1 (f) in the main text. The energy separation between the odd-parity state and its nearest even-parity state is $\Delta E\approx0.22$, matching the low-energy negative peak at $\omega\approx0.23$ in the postpump optical conductivity, as shown in Fig. 2(a) in the main text. However, $\Delta E$ becomes very small when $U$ is increased to $100$, as shown in Fig.~\ref{fig_S1} (b), and the odd- and even-parity states actually merge together when $U=+\infty$. In this scenario, the quarter-filled Peierls-Hubbard model transforms into the spinless noninteracting half-filled Su-Schrieffer-Heeger (SSH) model, which explains why we do not observe that low-frequency negative peak in the post-pump optical conductivity of the SSH model, see Fig. 4(a) in the main text.

On the other hand, we find that the energy separation between the odd- and even-parity states may be attributed to the finite-size effect even for finite $U$. For example, we present the results of $\langle j j \rangle$ and $\langle \tau \tau \rangle$ with $L=20$ and $U=10$ in Fig.~\ref{fig_S1}(c), where we find that $\Delta E$ is approximately $0.13$. In Fig~\ref{fig_S1}(d), we show the results of $\Delta E$ as a function of $1/L$ for $L=12$ and $L=20$, from which it can be inferred that $\Delta E$ may vanish in the thermodynamic limit ($L\rightarrow+\infty$).

We employ the Lanczos method for the calculations of these correlation functions, with the cutoff number $M=200$ and $300$ for $L=12$ and $L=20$, respectively.

%\bibliography{lt}


\end{document}